# How to deform an egg yolk?

## On the Study of Soft Matter Deformation in a Liquid Environment

Ji Lang,[1,2] Rungun Nathan,[1,3] Qianhong Wu,[1,2]*

[1]Cellular Biomechanics and Sports Science Laboratory, Villanova University, 800 E Lancaster Avenue, Villanova, PA 19085, USA.

[2]Department of Mechanical Engineering, Villanova University, 800 E Lancaster Avenue, Villanova, PA 19085, USA.

[3]Pennsylvania State University Berks, Reading, PA 19610, USA.

*To whom correspondence should be addressed: qianhong.wu@villanova.edu

**Abstract**

In this paper, we report a novel experimental and theoretical study to examine the response of a soft capsule bathed in a liquid environment to sudden external impacts. Taking an egg yolk as an example, we found that the soft matter is not sensitive to translational impacts, but is very sensitive to rotational, especially decelerating-rotational impacts, during which the centrifugal force and the shape of the membrane together play a critical role causing the deformation of the soft object. This finding, as the first study of its kind, reveals the fundamental physics behind the motion and deformation of a membrane-bound soft object, e.g., egg yolk, cells, soft brain matter, etc., in response to external impacts.

**1. Introduction**

Soft matter in a liquid environment widely exists in nature. Some examples include the soft brain matter that is bathed in the cerebrospinal fluid (CSF)[1–3] inside the hard skull, a soft egg yolk that is bathed in the fluidic egg white inside the eggshell, and red blood cells in our circulation system, etc. Deformability is one of the most important features of soft matter. For instance, the deformability is an indicator to decide whether red blood cells should be kept in the circulation system, or they should be cleared by the spleen[4,5].

Traumatic brain injury, on the other hand, is caused by large brain deformation as the head is exposed to sudden translational or rotational impact[6,7].

The deformation of soft matter in a fluid environment is a result of series of fluid-structure interactions. Different flow conditions and configurations could cause different types of deformation. Extensive studies have been performed to examine the deformation of soft capsules under shear flow[8–14], spinning flow[11], or through small channels[15–19]. In many applications, an active breaking of soft matters is needed. Hence, a significant amount of work has been done to examine the break-up mechanism of droplets and bubbles in a micro T-junction or a cross junction[20–23], in a pre-mix membrane[21], through a narrow constriction[24–26] or a permanent obstruction[27]. These studies demonstrate that large deformation of soft capsules in a liquid environment is induced by a fast change in the fluid field, such as the velocity gradient (e.g., the shear flow and the spinning flow), or velocity boundary conditions (e.g., sudden change of flow pathway).

For a soft matter bathed in a liquid environment and enclosed in a rigid container, an interesting yet fundamentally important question is that how one could damage or break the soft matter without breaking the container. This answer to this question could shed some light on the understanding of such problems as concussive brain injury. It also helps one to develop cell separation and processing equipment without affecting its viability. We did a simple experiment using a Golden Goose Egg Scrambler®, in which we used rotational forces to scramble an egg inside of the eggshell. It is puzzling to see how one could deform the yolk without cracking the shell. This experiment inspired us to study the fundamental physics governing the motion and deformation of soft matter in a liquid environment, using an egg yok as a sample system.

## 2. Experimental Setup

To damage or deform an egg yolk, one would try to shake and rotate the egg as fast as possible. Hence, two experimental setups, a translational impact setup, as shown in Fig. 1*A*, and a rotational impact setup, as shown in Fig. 1*B*, have been developed.

The egg yolk and the egg white used in the experiment are obtained from fresh eggs bought from a grocery store. As our goal is to break the egg yolk without breaking the eggshell, the eggshell is replaced by a transparent rigid container. For the translational impact, a 1.77kg hammer ((c) in Fig. 1A) falls freely along a vertical guide rail ((b) in Fig. 1A) from 1 m above the container to create the impact. At the bottom, a spring base ((d) in Fig. 1A) allows the container to move vertically. The acceleration of the container was measured by an accelerometer (Analog Devices, model ADXL 1004Z, with a bandwidth of 24,000 Hz, (e) in Fig. 1A). For the rotational impact setup, the container ((a) in Fig. 1B) is connected to an electric motor ((g) in Fig. 1B) that drives the container to rotate. The motion of the container is monitored and controlled by the motor controller. The motion and the deformation of the egg yolk are recorded through the transparent container by a high-speed camera (Phantom® Miro® C110) with a sample rate of 1,000 frames per second.

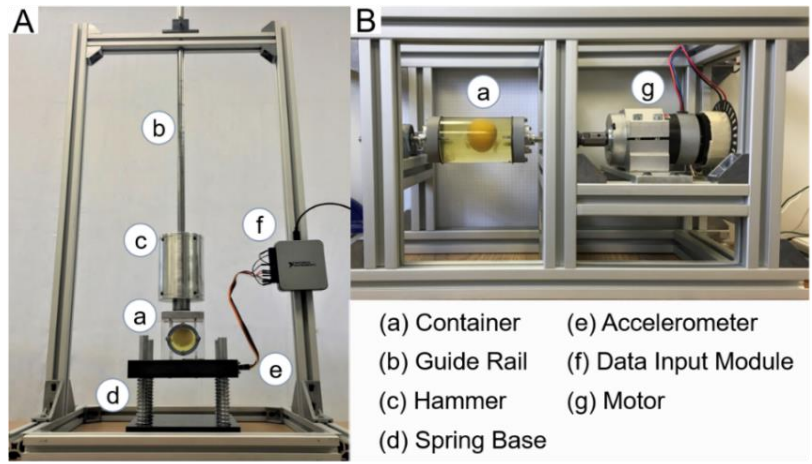

**Fig. 1.** (*A*) Translational impact setup. The hammer, (c), slides along the guide rail, (b), to impact the container, (a), which rests on the spring base, (d). The acceleration is measured by an accelerometer and the data is collected by the data input module. (*B*) Rotational impact setup. The electric motor, (g), drives the container, (a), to rotate.

## 3. Experimental Results

Firstly, the densities of the egg yolk and the egg white of six eggs have been measured. According to the measurements, the density of egg yolk is $1.045 \pm 0.020 \ g/ml$, while the density of egg white is $1.033 \pm 0.014 \ g/ml$. It confirms that the density of the egg yolk and egg white is very close to each other.

The egg yolk is enclosed by a thin, fragile, and soft membrane, the stretching of which is associated with the deformation of the egg yolk. Hence, it is crucial to evaluate the Young's modulus of the membrane in order to understand the stress it experiences during the deformation process. Egg yolk membrane was carefully peeled off from a fresh egg yolk and stored in a petri dish filled with water, as shown in Fig. 2A. A tensile test has been performed using a Psylothch µTS test system[28]. As the membrane is very soft and fragile, it was held on the water surface during the test. Five membrane specimens were stretched with a constant velocity of 0.1 $mm/s$, Fig 2B, until the membrane was broken, Fig. 2C. The tensile force and the stretched length were recorded. A Scanning Electron Microscope (SEM) was used to measure the thickness of the membrane, Fig. 2D, which indicates that the average thickness of the membrane is 3 $\mu m$. The width of the membrane sample is known, which, together with the thickness, provides the measurements of the cross-section area. Hence, the stress of the membrane is obtained. Fig. 2E shows the stress-strain relationship obtained in five testing cases. It shows that the membranes were linearly stretched before the maximum strain was reached. The membrane began to break as the strain was getting close to its maximum value. The slope of each curve provides the value of the Young's modulus, E= 2.64 ± 1.45 MPa, and the average strain when the break happened is 0.46 ± 0.14. It shows in Fig. 2D that the membrane thickness is not uniform, which might bring a relatively high variance in the measured physical properties.

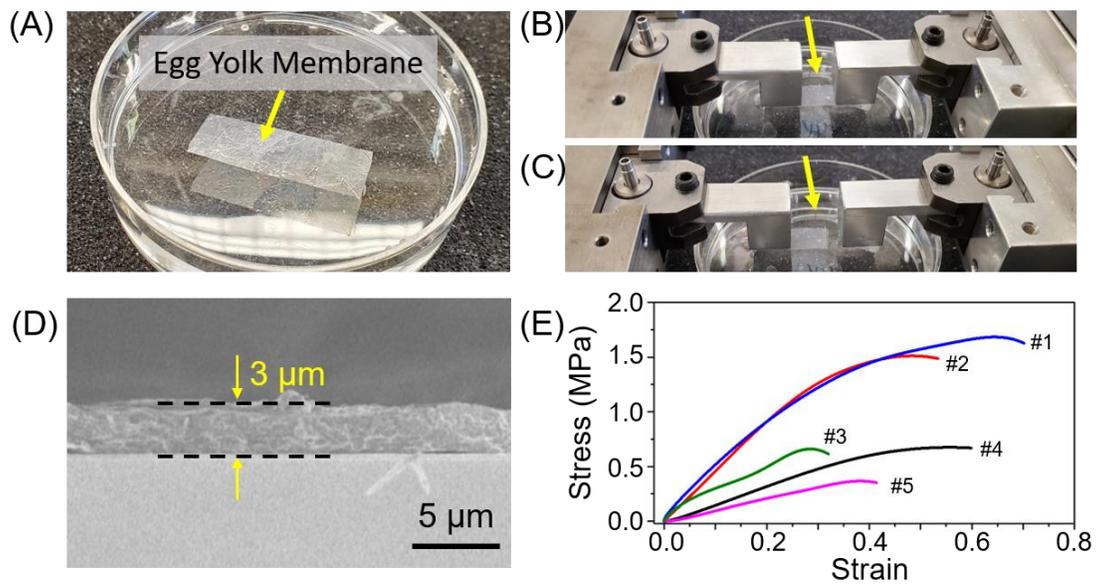

**Fig. 2.** (A) Egg yolk membrane specimen on water surface. (B) Specimen gripping. (C) Specimen fracture after testing. (D) SEM image of the cross-sectional view of egg yolk membrane. (E) strain-stress curves of five tested samples.

For the translational impact, the acceleration of the container was up to 600g, g=9.8m/$s^2$. As shown in Fig. 3A, the yolk had almost no deformation. This observation is quite surprising and counter-intuitive because one expects translational impacts would lead to the damage of the egg yolk. The reason behind this phenomenon is that the density difference between egg yolk and egg white is very small. Besides, both the yolk and the egg white are incompressible. Therefore, no relative motion was observed, and the whole container moved as a rigid body.

Two types of rotational impacts acceleration and deceleration were considered. For the accelerating-rotational impact, the container was set to rotate instantaneously from 0 rad/s to 400 rad/s within 1 s, after which it was kept at the constant angular velocity of 400 rad/s to establish steady-state. It shows in Fig. 3B that the yolk started from a spherical shape was subsequently compressed slightly in the radial direction at the center and hence stretched horizontally. Within 2 s, it was changed into an ellipsoid. Then, with a constant angular velocity, the egg yolk kept a stable shape as the one shown in the last panel of Fig. 2B for several minutes. The balance between shear stress, the centrifugal force, and the tension force in the membrane determines the slight deformation of the egg yolk.

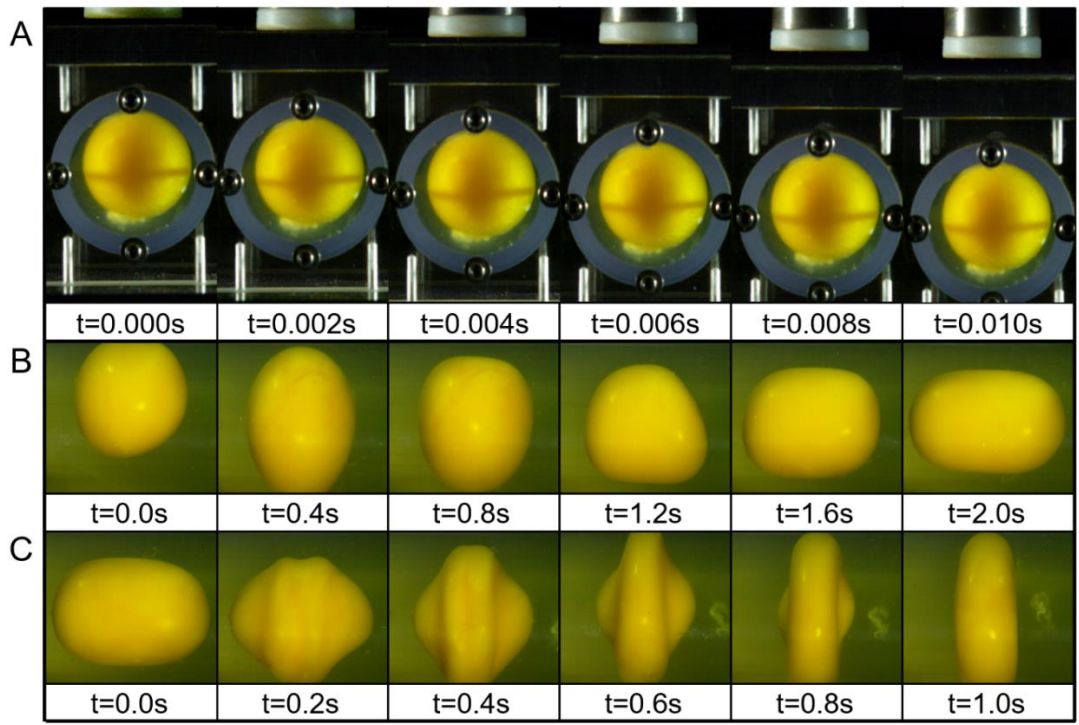

**Fig. 3** (*A*) Reactions of egg yolk under translational impact. the shell was impacted by a hammer to achieve the translational acceleration up to 600g, g=9.8m/s2. (*B*) Reactions of egg yolk under rotational acceleration impact. The container was set to rotate instantaneously from 0 rad/s to 400 rad/s within 1 s, after which it was kept at the constant angular velocity of 400 rad/s. (*C*) Reactions of egg yolk under rotational deceleration impact. The rotation speed of the container was reduced sharply from 400 rad/s to 0 rad/s within 1s to create a deceleration impact on the egg yolk.

The rotational speed of the container was then reduced sharply from 400 rad/s to 0 rad/s within 1s to create a deceleration impact on the egg yolk. The most intriguing result is obtained in the deceleration impact process, where the egg yolk experienced a significant deformation soon after the container started decelerating, which is shown in Fig. 3C. During the process, the yolk was tremendously squeezed horizontally and expanded radially in the center region. This large deformation obviously could cause severe damages to the yolk. At the end of the test, when the rotation is stopped, the deformed egg yolk slowly resumed its original spherical shape (i.e., Fig. 3B at t = 0.0 s), which takes approximately one minute. Such an unexpected observation inspires us to investigate the mechanism behind it further. The video of the translational impact case, the rotational acceleration case, and the rotational deceleration case can be found in the supplemental information.

To quantify the membrane stretch during the experiments, we estimate the averaged strain and stress of the yolk membrane. In the beginning, the yolk was in a spherical shape, as shown in Fig. 3B at t = 0.0 s. Its strain was zero. When the yolk began to deform, its total volume remained unchanged due to incompressibility. Hence, the surface area of the membrane would change during the acceleration/deceleration process. The change of the total surface area with respect to the initial surface area is the averaged strain of the membrane. Because the yolk established an axisymmetric shape as it picked up the rotational speed during the acceleration phase and remained to be axisymmetric during the deceleration phase, one could obtain the total surface area of the yolk at different instants by surface integration. Fig. 4A shows the coordinate system and a representative cross-section of the yolk obtained from Fig 3. The geometry is axisymmetric about the x-axis. The highlighted region represents an infinitesimal element, which is treated as a frustum. The surface area of this infinitesimal element, $dS$, is

$$dS = [\pi y + \pi(y + dy)]\sqrt{dx^2 + dy^2}. \tag{1}$$

By integrating, the total surface area, $S(t)$, a function of time, could be obtained. Then, the averaged strain of the membrane, defined as $\epsilon = [S(t) - S_0]/S_0$, is obtained. Here $S_0$ is the initial surface area of the yolk, i.e., Fig. 3B at t = 0.0 s.

The change of the averaged strain and stress during the deceleration process is shown in Fig. 4B. The y-axis on the left represents the strain, while the y-axis on the right is the averaged stress obtained from multiplying the strain with the Young's modulus. It shows in Fig. 4B that the strain and stress decreased first as the yolk was squeezed in the axial direction into a more spherical shape, corresponding to Fig. 3C, t =0.2 s. Then, as the yolk was deformed more seriously in the axial direction and deviated from the spherical shape, the strain and the stress increased rapidly, corresponding to Fig. 3C, t =0.4 s and t = 0.6 s. With the further decrease of the angular velocity, the yolk tends to restore its original shape. Therefore, the strain and stress decreased again. The maximum strain is 0.25. According to the tensile test shown in Fig. 2E, the membrane is still in the elastic region. The maximum stress in the membrane, σmax= 0.66 MPa, is then obtained for the deceleration process. Since the deformation is relatively small during the acceleration phase, we focus our discussion of averaged strain and stress in the deceleration process.

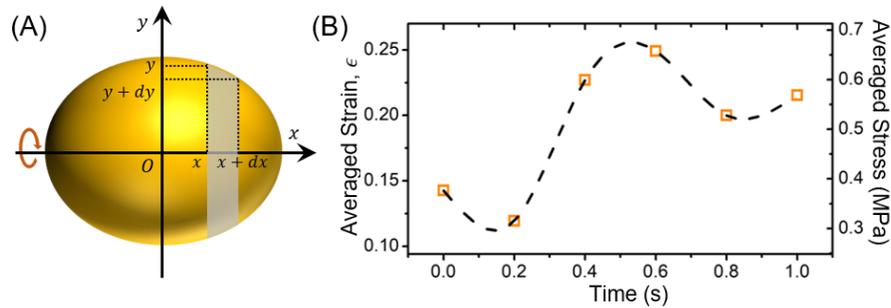

**Fig. 4** (*A*) Schematic of surface area integration. (*B*) Development of averaged strain & stress with time for deceleration case.

**4. Theoretical Study**

To understand the fundamental physics causing the large deformation of the egg yolk in the deceleration impact process, we would like, as a first try, to examine the response of the the velocity and the pressure field when the motion of the outer shell is rapidly altered. As shown in Fig. 5A, an impermeable spherical membrane is placed at the center of a cylindrical container that is filled with liquid. The container can rotate with angular velocity $\pm \omega_0$. The membrane is very thin, and thus there is no pressure gradient across it. It divides the container into Region I and Region II. As an approximation, everywhere on the membrane shares the same angular velocity, which is determined by the torque caused by the shear stress on both sides of the membrane. A cylindrical coordinate system (r, θ, z) is shown in Fig. 5A. The container has a radius of $R_o$ and a length of $2L$. The spherical membrane has a radius of $R_i$. Because the problem is symmetric about r axis and axisymmetric about the z-axis, only the first quarter is plotted.

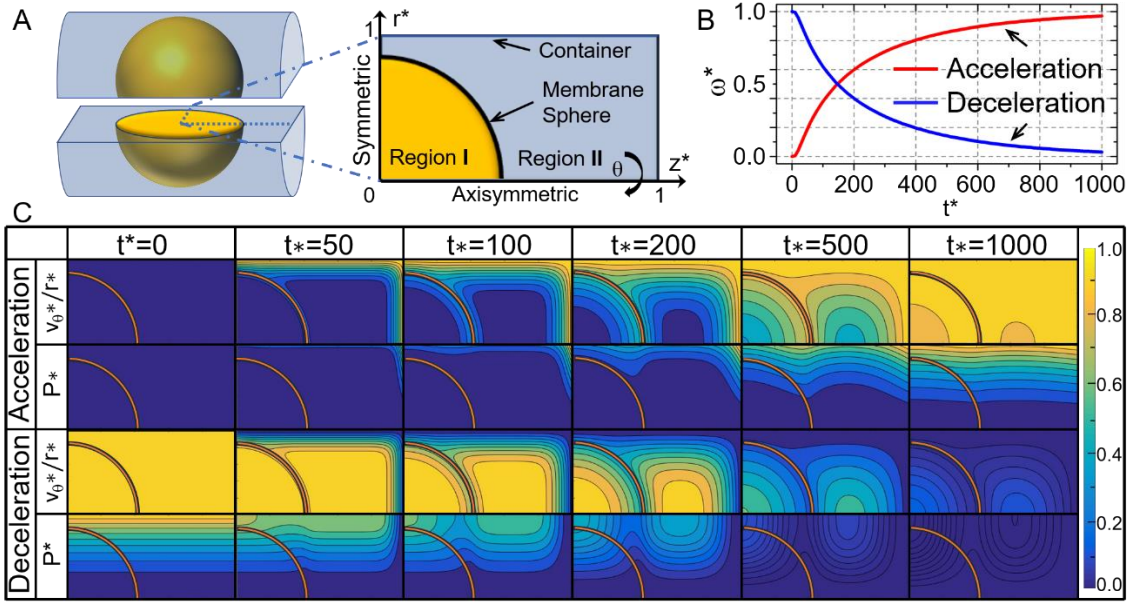

**Fig. 5**. (A) Sketch of the 2-D rotating model. The space inside the container was divided as Region I, which is enclosed by the membrane, and Region II, which is between the membrane and the container. The change of the velocity of the container would first penetrate into viscous gap between the membrane and the container, and then influence the velocity of the membrane as well as the fluid inside the membrane. (B) $\omega^*$ under the rotational acceleration impact and rotational deceleration impact. (C) Distribution of the angular velocity, $v_\theta^*/r^*$, and pressure field, $P^*$, under the acceleration impact and the deceleration impact.

It is reasonable to assume the fluid velocity in the radial and axial directions are much smaller than that in the circumferential direction. Also, because we are mainly interested in the mechanism that causes the deformation of the egg yolk, we chose a small-time scale during which the shape of the membrane has no time to change. The Navier-Stokes equation for the fluid flow is then simplified as

$$2\frac{v_\theta^{*2}}{r^*} = \frac{\partial P^*}{\partial r^*}, \quad \frac{\partial v_\theta^*}{\partial t^*} = \frac{\text{Sr}}{\text{Re}}\left[\frac{\partial}{\partial r^*}\left(\frac{1}{r^*}\frac{\partial(r^* v_\theta^*)}{\partial r^*}\right) + \frac{R_0^2}{L^2}\frac{\partial^2 v_\theta^*}{\partial z^{*2}}\right], \tag{2a, b}$$

where $v_\theta^* = \frac{v_\theta}{V_0}$, $P^* = \frac{2P}{V_0^2 \rho}$, $r^* = \frac{r}{R_o}$, $z^* = \frac{z}{L}$, $t^* = \frac{t}{t_0}$, $V_0 = r\omega_0$, $\nu$ is the kinematic viscosity of the fluid, Reynolds number, $\text{Re} = V_0 R_0/\nu$, and Strouhal number, $\text{Sr} = t_0 V_0/R_0$. Here $v_\theta(r,z,t)$ is the circumferential velocity, $P$ is the pressure, $\rho$ is the fluid density, and $t_0 = \frac{2\pi}{\omega_0}$, is the characteristic time.

The rotational acceleration of the membrane is determined by the shear stress on both sides of the surface,

$$\frac{d\omega^*}{dt^*} = \frac{2\pi R_o^3 \rho v t_p}{I}\left(\frac{R_o}{L}\int_A \frac{dv_\theta^*}{dz^*} r^{*2} dr^* + \frac{L}{R_o}\int_A \frac{dv_\theta^*}{dr^*} r^{*2} dz^*\right), \tag{3}$$

where $\omega^* = \omega/\omega_0$, $\omega_0$ is the angular velocity of the membrane, $I$ is the inertia of the membrane, and A represents the area of the membrane's both sides.

For the rotational acceleration case, the initial condition requires that when $t^* = 0$, $v_\theta^* = 0$ and $\omega^* = 0$. While the boundary conditions state that, $v_\theta^* = 1$ on $r^* = 1$ or $z^* = 1$, and $v_\theta^* = r^*\omega^*$ on the membrane.

For the rotational deceleration case, the initial condition is given as when $t^* = 0$, $v_\theta^* = 1$ and $\omega^* = 1$. And the boundary conditions require that, $v_\theta^* = 0$ on $r^* = 1$ or $z^* = 1$, and $v_\theta^* = r^*\omega^*$ on the membrane.

Eqns. (2) and (3) have been solved numerically. The solutions of $\omega^*$ during the acceleration and deceleration process are given in Fig. 5B. The solutions of the angular velocity, $v_\theta^*/r^*$, and pressure response, $P^*$, at six representative moments during the acceleration and deceleration process, are given in Fig. 5C. When the outer cylinder starts to rotate, the momentum is penetrated through the viscous fluid gap to the membrane-bound ball, then to the fluid inside the ball. Thus, at the same axial position, $z^*$, the angular velocity, $v_\theta^*/r^*$, of the fluid near the inner side of the membrane is smaller than that of fluid near the outer side of the membrane. Hence, the pressure outside the membrane is larger than the centrifugal force of the fluid that is enclosed inside the membrane. This pressure difference will cause the compression of the yolk at the center, leading to the one as observed in Fig. 3B.

Once the steady state is achieved, the outer shell is suddenly brought to rest while the fluid inside the shell is still moving. This makes the angular velocity of on the inner side of the membrane is larger than that of the fluid near the outer surface of the membrane, making the outside pressure is far less than the centrifugal force of the fluid enclosed inside the membrane. Due to its extremely soft nature, the yolk is not able to hold its shape. Hence, the yolk will expand and squeeze out the fluid in the gap between the yolk and the shell, leading to the one observed in Fig. 3C.

This problem involves the inertia force (or centrifugal force) and the shear stress penetration, therefore, the Re is important. Besides, as shown in Eqn. (2), the Sr describes the transient process caused by the sudden change in the shell velocity. A smaller $\frac{Sr}{Re}$ will decrease the time-dependent effect and the inertia force and slow down the shear stress penetration. Physically, it suggests a slower boundary change or a smaller viscosity. Hence, the yolk deformation will decrease correspondingly. Another important parameter is $R/R_0$ ($R$ is the radius of the yolk at different axial locations), which describes the local gap height for the momentum to transfer from the outer cylinder to the inner object. $R/R_0$ takes different values at different axial locations. Hence, the time for the sudden change in the velocity to be felt by the membrane is different. If $\frac{\partial R/R_0}{\partial z} = 0$, the deformation would not happen.

## 5. Conclusion

The experimental and theoretical study presented herein shows that the soft egg yolk is not sensitive to translational impacts, but is very sensitive to rotational, especially deceleration rotational impacts, during which the centrifugal force plays a critical role. This finding provides a new perspective for understanding the response of a membrane-bound soft object in the liquid environment to sudden external impacts. It is noticed that the phenomena discussed in this paper seem to be similar to a spinning droplet in tensiometer undergoing rotation[29–32]. However, the physical mechanisms behind these two phenomena are different. The spinning droplet in the tensiometer is usually analyzed in a static state, which requires a density difference between the droplet and the surrounding fluid. The egg deformation is a dynamic process of which the density difference between the yolk and the egg white is very small. For the droplet spinning problem, the essential physics is the balance between the surface tension and the centrifugal force, while for this study, the centrifugal force is mainly balanced by the pressure force caused by fluid flow. The existence of the membrane separates the two fluids. The momentum transfer from the outer cylinder to the membrane and then to the liquid yolk inside. Since the travel distance is different, the pressure field of the fluid is disturbed, leading to the deformation of the membrane-bound egg yolk and the fluid flow outside of the membrane. The role of the tension force in the membrane is to force the membrane to rotate at a constant angular velocity and hold the soft matter together.

## 6. Supplemental Information

Movie 1. Reaction of egg yolk under translational impact.

Movie 2. Reaction of egg yolk under rotational acceleration impact.

Movie 3. Reaction of egg yolk under rotational deceleration impact.

**7. Acknowledgments**

**Funding**: This work is supported by National Science Foundation CBET Fluid Dynamics Program under Award #1511096. The authors would like to thank Mr. Dong Zhou for his help in measuring Yong's modulus of the yolk membrane.

**8. Author Contributions**

Q. W. and J. L. conceived of the presented idea, designed the experiment, developed the theoretical model, performed data analysis, and drafted the manuscript. J. L. and R. N constructed the experimental setup and performed the experiments. J. L. developed theoretical model. Q. W. approved the final manuscript.

**9. References**

[1]    Tawse K 2008 Cerbrospinal Fluid-tissue Interactions in the Human Brain *J. Young Investig.*

[2]    Linge S O, Haughton V, Løvgren A E, Mardal K A and Langtangen H P 2010 CSF flow dynamics at the craniovertebral junction studied with an idealized model of the subarachnoid space and computational flow analysis *Am. J. Neuroradiol.* **31** 185–92

[3]    Kertzscher U, Schneider T, Goubergrits L, Affeld K, Hänggi D and Spuler A 2012 In vitro study of cerebrospinal fluid dynamics in a shaken basal cistern after experimental subarachnoid hemorrhage *PLoS One* **7** e41677

[4]    Pivkin I V., Peng Z, Karniadakis G E, Buffet P A, Dao M and Suresh S 2016 Biomechanics of red blood cells in human spleen and consequences for physiology and disease *Proc. Natl. Acad. Sci.* **113** 7804–9

[5]    Kim J, Lee H and Shin S 2015 Advances in the measurement of red blood cell deformability: A brief review *J. Cell. Biotechnol.* **1** 63–79

[6]    CDC. 1997 Sports-related recurrent brain injuries—United States *Morb. Mortal. Wkly. Rep.* **46** 224–7

[7]    Thurman D J, Branche C M and Sniezek J E 1998 The epidemiology of sports-related traumatic brain injuries in the United States: recent developments. *J. Head Trauma Rehabil.* **13** 1–8

[8]    Barthès-Biesel D, Diaz A and Dhenin E 2002 Effect of constitutive laws for two-dimensional membranes on flow-induced capsule deformation *J. Fluid Mech.* **460** 211–22

[9]     Lac E and Barthès-Biesel D 2005 Deformation of a capsule in simple shear flow: Effect of membrane prestress *Phys. Fluids* **17** 072105

[10]    Chang K S and Olbricht W L 1993 Experimental studies of the deformation of a synthetic capsule in extensional flow *J. Fluid Mech.* **250** 587–608

[11]    Pieper G, Rehage H and Barthès-Biesel D 1998 Deformation of a Capsule in a Spinning Drop Apparatus *J. Colloid Interface Sci.* **202** 293–300

[12]    Foessel E, Walter J, Salsac A V. and Barthès-Biesel D 2011 Influence of internal viscosity on the large deformation and buckling of a spherical capsule in a simple shear flow *J. Fluid Mech.* **672** 477–86

[13]    Rahmat A, Barigou M and Alexiadis A 2019 Deformation and rupture of compound cells under shear: A discrete multiphysics study *Phys. Fluids* **31** 051903

[14]    Hassan M R, Zhang J and Wang C 2018 Deformation of a ferrofluid droplet in simple shear flows under uniform magnetic fields *Phys. Fluids* **30** 092002

[15]    Secomb T W, Styp-Rekowska B and Pries A R 2007 Two-dimensional simulation of red blood cell deformation and lateral migration in microvessels *Ann. Biomed. Eng.* **35** 755–65

[16]    Quinn D J, Pivkin I, Wong S Y, Chiam K H, Dao M, Karniadakis G E and Suresh S 2011 Combined simulation and experimental study of large deformation of red blood cells in microfluidic systems *Ann. Biomed. Eng.* **39** 1041–50

[17]    Che Z, Yap Y F and Wang T 2018 Flow structure of compound droplets moving in microchannels *Phys. Fluids* **30** 012114

[18]    Luo Z Y and Bai B F 2019 Solute release from an elastic capsule flowing through a microfluidic channel constriction *Phys. Fluids* **31** 121902

[19]    Ye T and Peng L 2019 Motion, deformation, and aggregation of multiple red blood cells in three-dimensional microvessel bifurcations *Phys. Fluids* **31** 021903

[20]    Garstecki P, Fuerstman M J, Stone H A and Whitesides G M 2006 Formation of droplets and bubbles in a microfluidic T-junction—scaling and mechanism of break-up *Lab Chip* **6** 437

[21]    Tan J, Xu J H, Li S W and Luo G S 2008 Drop dispenser in a cross-junction microfluidic device:

Scaling and mechanism of break-up *Chem. Eng. J.* **136** 306–11

[22]    Link D R, Anna S L, Weitz D A and Stone H A 2004 Geometrically Mediated Breakup of Drops in Microfluidic Devices *Phys. Rev. Lett.* **92** 4

[23]    Xu J H, Li S W, Tan J and Luo G S 2008 Correlations of droplet formation in T-junction microfluidic devices: From squeezing to dripping *Microfluid. Nanofluidics* **5** 711–7

[24]    Rosenfeld L, Fan L, Chen Y, Swoboda R and Tang S K Y 2014 Break-up of droplets in a concentrated emulsion flowing through a narrow constriction ed G Gompper and M Schick *Soft Matter* **10** 421–30

[25]    Le Goff A, Kaoui B, Kurzawa G, Haszon B and Salsac A V 2017 Squeezing bio-capsules into a constriction: Deformation till break-up *Soft Matter* **13** 7644–8

[26]    Tryggvason G 2020 The passage of a bubble or a drop past an obstruction in a channel *Phys. Fluids* **32** 023303

[27]    Wang X, Zhu C, Fu T and Ma Y 2015 Bubble breakup with permanent obstruction in an asymmetric microfluidic T-junction *AIChE J.* **61** 1081–91

[28]    Kim J-H, Nizami A, Hwangbo Y, Jang B, Lee H-J, Woo C-S, Hyun S and Kim T-S 2013 Tensile testing of ultra-thin films on water surface *Nat. Commun.* **4** 2520

[29]    Langbehn B, Sander K, Ovcharenko Y, Peltz C, Clark A, Coreno M, Cucini R, Drabbels M, Finetti P, Di Fraia M, Giannessi L, Grazioli C, Iablonskyi D, LaForge A C, Nishiyama T, Oliver Álvarez de Lara V, Piseri P, Plekan O, Ueda K, Zimmermann J, Prince K C, Stienkemeier F, Callegari C, Fennel T, Rupp D and Möller T 2018 Three-Dimensional Shapes of Spinning Helium Nanodroplets *Phys. Rev. Lett.* **121** 255301

[30]    Vorobev A and Boghi A 2016 Phase-field modelling of a miscible system in spinning droplet tensiometer *J. Colloid Interface Sci.* **482** 193–204

[31]    Jiang Y, Zhao C, Cheng T and Zhou G 2019 Theoretical model in cylindrical coordinates to describe dynamic interfacial tension determination with spinning drop tensiometry *Chem. Phys.* **525** 110409

[32]    Ullmann K, Poggemann L, Nirschl H and Leneweit G 2020 Adsorption process for phospholipids

of different chain lengths at a fluorocarbon/water interface studied by Du Noüy ring and spinning drop *Colloid Polym. Sci.* **298** 407–17